# Imaging Electron Motion in Graphene

**Sagar Bhandari**[1] and **Robert M. Westervelt**[1,2]
School of Engineering and Applied Sciences, Harvard University, Cambridge, MA 02138, U.S.A
Department of Physics, Harvard University, Cambridge, MA 02138, U.S.A

Email: sbhandar@fas.harvard.edu

**Abstract.** Research in semiconductor physics has advanced to the study of two-dimensional (2D) materials where the surface controls electronic transport. A scanning probe microscope (SPM) is an ideal tool to image electronic motion in these devices by using the SPM tip as a scanning gate. In prior work for a two dimensional electron gas (2DEG) in a GaAs/AlGaAs heterostructure, a number of phenomena were imaged, including electron flow from a quantum point contact, and tuning of a few electron quantum dot. This approach was also used to study InAs quantum dots grown in a InAs/InP nanowire heterostructure. New two-dimensional materials such as graphene show great promise for fundamental research and applications. We have extended or prior work to image the motion of electrons along cyclotron orbits in single atomic layer graphene passing from one point contact to a second point contact on the first magnetic focusing peak in graphene. The charged SPM tip defects electrons passing from one contact to the other, changing the conductance. A plot of the change in conductance vs. tip position presents an image of electron flow. A low temperature Scanning Capacitance Microscope (SCM) with a sensitive charge preamplifier located near the SPM tip achieves a charge noise level 0.13 e/Hz$^{1/2}$ with high spatial resolution 100 nm, which promises to be useful to study electronic behavior in nanoscale devices.

**Introduction**

A cooled scanning probe microscope (SPM) has been proven to be a powerful tool to image electron states in a two dimensional electron gas (2DEG) inside a GaAs/Al$_3$Ga$_7$As heterostructure [1-14]. Quantum hall liquid edge states in a 2DEG were imaged using scanning single electron transistor (SET) as a charge sensor [6]. By using the capacitively coupled SPM tip to deflect electron trajectories, coherent electron flow from a quantum point contact (QPC) was imaged by displaying the change in conduction as the tip was raster scanned above the sample [7-14]. Magnetic focusing of electrons in a 2DEG was imaged using this technique [13,14]. A charged SPM tip can also be used to capacitively tune a quantum dot. by displaying the dot conductance as the tip is raster scanned above; creating bullseye pattern in which the rings of high conductance correspond to Coulomb-blockade conductance peaks [15-18]. This approach as been applied to quantum dots formed in a GaAs 2DEG by surface gates, and to an InAs dot grown in an InAs/InP/InAs nanowire heterostructure [15-18]. Quantum dots in single walled carbon nanotubes were also imaged using Scanning Force Microscopy [20-21]. Following the discovery of 2D materials such as graphene, the SPM has been proven to be an invaluable tool to shed light on the new physics of these systems. Electron hole puddles in graphene were imaged using scanning Single Electron Transistor (SET) [22] and STM techniques [23]. Using a capacitively coupled SPM tip, universal conductance fluctuations in graphene was spatially mapped for electrons in graphene [24, 25]. Quantum dots in graphene nanostructures were imaged to reveal Coulomb-blockade rings [26].

A wealth of 2D materials have been discovered in recent years with exotic electronic properties [27]. Electrons in graphene have energies that increase linearly with their momentum at low energies; they are massless, chiral, Dirac fermions described by the Dirac equation [28] that display new phenomena, including the Anomalous Quantum Hall effect, Klein tunneling, and Berry's phase [29-37]. Because graphene is two dimensional material with it's surface completely exposed to the environment, the interface to the substrate plays an important role in its electronics properties. Graphene devices on a silicon substrate typically have a mobility ~ 10,000 cm$^2$/Vs. Freely suspending graphene drastically improves the mobility to ~200,000 cm$^2$/Vs, but limits the device architecture and functionality. Encasing the graphene in two hexagonal boron nitride (hBN) layers greatly enhances the mobility [38-40]. As a result, electrons can travel several microns without scattering at low temperatures and follow classical trajectories as their motion becomes ballistic. As shown below, we have used a cooled SPM to image cyclotron orbits of electrons in a hBN/graphene/hBN device in a magnetic field showing ballistic transport [41, 42].

**Methods**

Scanning probe microscopy involves using a sub-nanometer sized tip as a local probe to study the electrons in materials at nanoscale. In a scanning gate measurement at low temperatures, local trajectories of electrons can be mapped in the sample by having a conducting tip directly above it raster scan the sample while simultaneously measuring the conductance change through the sample. A charge on the tip creates an image charge and a corresponding density change in the 2DEG below, shown by simulations in Fig. 1(a). Scattering by the image deflects electron paths and changes the conductance of the sample, so a map of the conductance *vs*. tip position reveals the electronic trajectories in the sample.

Using this technique, we have imaged the coherent flow of electrons from a quantum point contact in GaAs/AlGaAs heterostructure [7]. Fig. 1(b) shows an image of electron flow recorded using a scanning gate microscope. As shown in Fig. 1(c) the tip voltage depletes a small divot in the electron gas below, that scatters electron waves back through the QPC, reducing its conductance. By displaying the QPC conductance as the tip is raster scanned across the sample, an image of electron flow is obtained. In Fig. 1 (d) the maps of electron trajectories revealed fringes corresponding to the interference of electron waves. The fringes were spaced at half a Fermi wavelength confirming the existence of coherent waves of electrons in 2DEG.

Fig. 2(a) shows the scanning gate setup to image electrons confined in quantum dots in a nanowire. The tip is scanned at a constant height above the nanowire while measuring the conductance through the device. The tip is used as a movable gate to locate the quantum dot and tune its charge down to first electron. The quantum dot is formed in a InAs/InP heterostructure by having two InP barriers as shown in Fig. 2(b). The diameter of the wire is 50 nm with a length of 3 $\mu$m. The quantum dot size is 18 nm long formed between the 8 nm thick InP barriers. The images show concentric rings of high conductance around the dot corresponding to coulomb blockade peaks as single electron is added into the dot. Only one ring was observed for the first electron added to the dot as shown in Fig. 2 (c).

The cooled scanning probe microscope (SPM) used to image electron motion inside graphene devices is described here [42]. The microscope consists of a head assembly containing the tip, and a cage containing the scanning piezotube and the sample holder, shown in Fig. 3. The sample stage is attached to the top of the piezotube. A typical scan is performed by translating the piezotube horizontally below the tip after setting the SPM tip height above the sample. A coarse positioning system allows the upper head to be moved horizontally relative to the lower head through the vertical motion of two wedges by screws that go to the top of the Dewar.

The cooled SPM uses a piezoresistive cantilever, in which the cantilever deflection is measured by a bridge circuit. High voltages lines that control the piezotube are shielded by stainless steel coaxial cables to reduce the pickup in the sample measurement leads. The SPM electronics include the force feedback controller to measure the height of the tip above the sample, an xy voltage controller for scanning controlled by a computer equipped with a DAC to generate the required voltages. A program integrates SPM sweeps of the sample with conductance data acquisition to create images of electron flow. The microscope lies inside a liquid Helium Dewar that can be operated between 4.2K and 1.5K.

Measurements of the capacitance between the SPM tip and a sample provide the opportunity for CV profiling at the nanoscale. We have also developed an SPM to image the tip-to-sample capacitance $C_{tip}$ vs. dc tip voltage, motivated by the work of Ashoori [1-6, 42, 43]. A cooled charge amplifier, shown in Fig. 4(a), located next to the tip is used to measure the capacitance. The input noise measured at 4.2 K is 20nV/√Hz. and the charge noise is 0.13e$^-$/√Hz. The capacitance-profiling setup was tested on a sample consisting of a 15 nm thick gold film on a SiO$_x$/Si substrate, shown in Fig. 4(b). The tip to sample capacitance measured as it was scanned across the edge of the gold film, is shown in Fig. 4(c). A change in capacitance of 30 aF was measured as the tip was scanned across the edge of the gold film, with a spatial resolution of 100 nm. This tool allows one to probe a quantum dot with just two electrodes, the SPM tip and the sample contact, facilitate further understanding of low dimensional nanoscale systems.

**Imaging Cyclotron Orbits in Graphene**

In the following section we present recent results imaging cyclotron orbits in graphene in the magnetic focusing regime [41, 42]. Magnetic focusing occurs when electrons travelling from one point contact to another in a perpendicular magnetic field B when the contact spacing L equals the diameter of a cyclotron orbit, as shown in Fig. 5(a). Magnetic focusing increases the chemical potential in the receiving contact, due to the high flow of incoming electrons. As the B continues to increase, additional peaks can occur when the L is an integer multiple of the cyclotron diameter, if the electron orbit skips along the edge [44-46].

To image cyclotron orbits, we used a hBN-graphene-hBN device etched into a hall-bar geometry [41, 42] with two broad contact at the ends and two narrow contacts along each side, separated by 2.0 µm. The degree of focusing is measured by the transmission of electrons between two point contacts along one edge, determined through the voltage measured at the receiving contact and the change in transresistance DR$_m$. Our cooled scanning gate microscope images image cyclotron trajectories in graphene. The tip creates a local dip in charge density that deflects electron trajectories and changes the transmission. As shown in Fig. 5(b), the cyclotron trajectories deflect from original path creating a drop in voltage measured at receiving contact. The inset shows the simulated image of transmission change of electrons trajectories as the tip is raster scanned above the sample- red represents the drop while blue is increase in transmission. This technique allows us to map the cyclotron orbits of electrons in graphene.

Magnetic focusing transmission peaks in the transmission between point contacts are shown in Fig. 6(a), which displays the transresistance (R$_m$) vs. n$^{1/2}$ and B at T = 4.2 K. The first magnetic focusing peak is clearly shown as the red band of decreased transmission between the two contacts that occurs when the cycltron diameter is equal to the contact spacing. In addition, one sees signs of a second magnetic focusing peak at twice the magnetic field, although the signal is not strong, presumably due to scattering at the sample edge.

Imaging results are shown in Fig. 6(b). In the absence of a magnetic field, in Fig. 6(b), no image is seen, as expected. However, when on applies a perpendicular magnetic field on the first magnetic focusing peak in Fig. 6(c) a clear image of the cyclotron orbit is obtained. The width and shape are due to fact that electron trajectories enter the sample at all angles, as illustrated by Fig. 5 above. The blue area of increased transmission when the tip is near the edge of the sample results from the tip deflecting trajectories away from the diffusely scattering edge of the graphene sample, as discussed below.

Figures 7(a) and (b) show tiled plots of the experimental SPM images and ray tracing simulations of electron transmission vs. B and n over the region of the first magnetic focusing peak shown in Fig. 6(a). Clear visualization of the cyclotron orbit is shown, both in the SPM images and the ray-tracing simulations, which closely correspond. These results demonstrate that the cooled SPM can image ballistic orbits of electron in graphene, opening the way to the development of devices based on ballistic transport.

In addition to the red cyclotron orbits in Fig. 7, we also see blue regions of increased transmission between contacts. At low magnetic fields B, the blue region is far from the edge, while at high B it appears very close to the edge of the sample. This observation suggests that the tip enhances transmission by deflecting electrons into the receiving contact. At low B, cyclotron diameter is large compared with the contact spacing, and the electrons do not make it to the receiving contact. As shown in Fig. 8(a) the tip can deflect some of these trajectories to the receiving contact, increasing the transmission. At high B, the cyclotron orbit is shorter than the contact separation, and electrons diffusively scatter off the edge, instead of reaching the second contact. In this case the tip can bounce electrons away from the edge, into the receiving contact, as shown in Fig. 8(b).

**Conclusion**

Imaging provides us with ability to locally probe electronic motion in nanoscale structures. The growing number of two dimensional materials such as graphene and transition metal dichalcogenides such as $MoS_2$, $WSe_2$ and BP, 2D topological insulators offers a great opportunity to probe novel physics in low dimensional systems [47, 48]. The capabilities of cooled scanning gate microscope are improving. In our group, improvements allow coarse positioning of tip and sample at helium temperature which can be critical for its operation. A technique to permit local CV profiling at nanoscale resolution was also demonstrated. Using the cooled SPM, we were able to image cyclotron orbits of electrons in graphene. The B and n dependence of the cyclotron orbits agree well with simulations of electron trajectories.

**Acknowledgements**

The SPM imaging research and the ray-tracing simulations were supported by the U.S. DOE Office of Basic Energy Sciences, Materials Sciences and Engineering Division, under grant DE-FG02-07ER46422. Nanofabrication was performed in part at the Center for Nanoscale Systems (CNS), a member of the National Nanotechnology Coordinated Infrastructure Network (NNCI), which is supported by the National Science Foundation under NSF award no. 1541959. CNS is part of Harvard University.

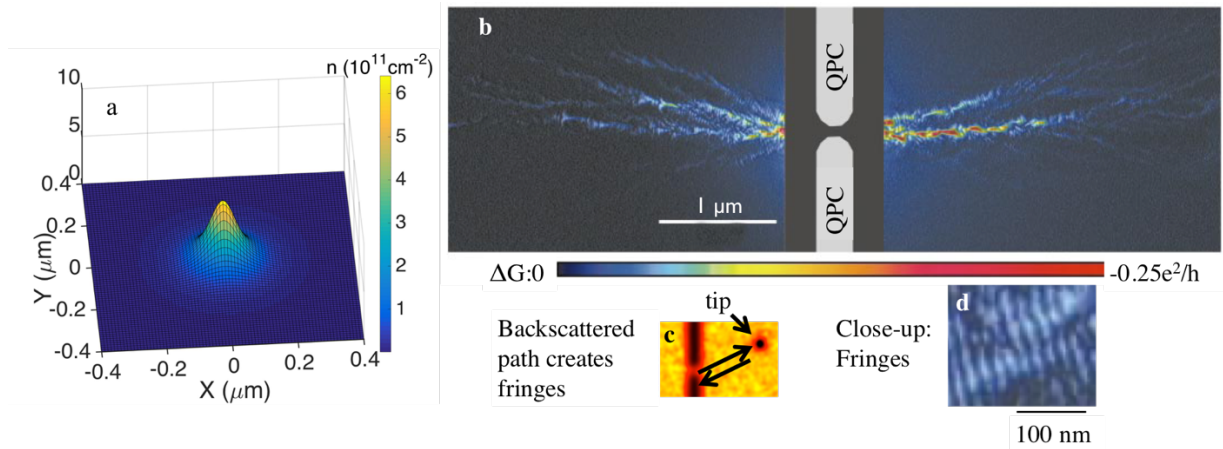

Figure 1: (a) Charge density profile created by a charged tip above the sample surface. The tip voltage depletes a small divot in the electron gas below that scatters electron waves back through the QPC, reducing its conductance, as shown in the inset. By displaying the QPC conductance as the tip is raster scanned across the sample, an image of the electron flow is obtained. (b) Scanning probe microscopy images at temperature 4.2K of electron flow from a quantum point contact. Branches in flow are created by scattering from Si donor ions. The images reveal interference fringes spaced by half the Fermi wavelength, shown in the inset, which verified that the electron waves are coherent [7].

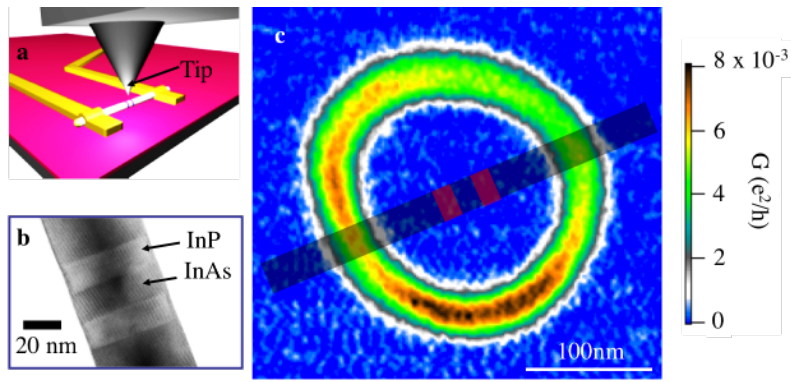

Figure 2: Tuning the charge on a single electron in a quantum dot in a InP/InAs/InP nanowire. (a) Scanning Gate Microscope setup (b) TEM image of an InP/InAs/InP quantum dot formed by two InP barriers (c) Image of Coulomb blockade conductance *vs.* tip position for the last electron in the quantum dot at temperature 4.2K [17].

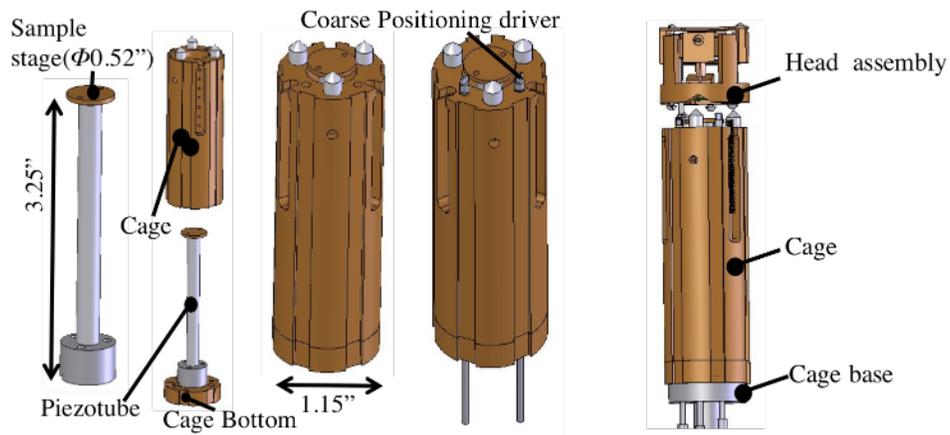

Figure 3: The leftmost figure shows the piezotube with sample stage fixed to a macor insulator on the bottom. Placing the piezotube onto the cage makes the cage assembly shown in the next figure. The rightmost figure shows the coarse positioning devices used to position the SPM cantilever and tip above the sample stage [42].

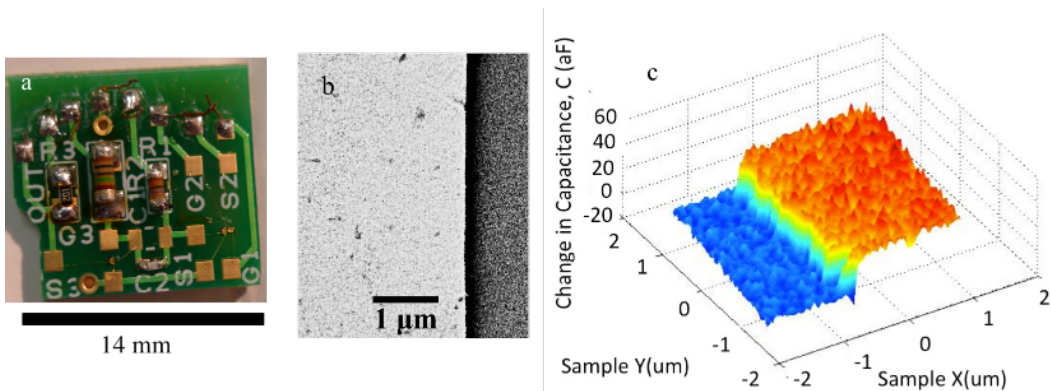

Figure 4: Scanning capacitance microscopy (a) A charge preamplifier mounted next to the SPM cantilever. (b) Scanning electron microscope image of a gold-film sample. (c) Capacitance-scan of a 1 um diameter tungsten tip above the sample, showing the change in capacitance as the tip is scanned at a height of 20 nm above the edge of the gold film. The FWHM of the capacitance transition is 100 nm [42, 43].

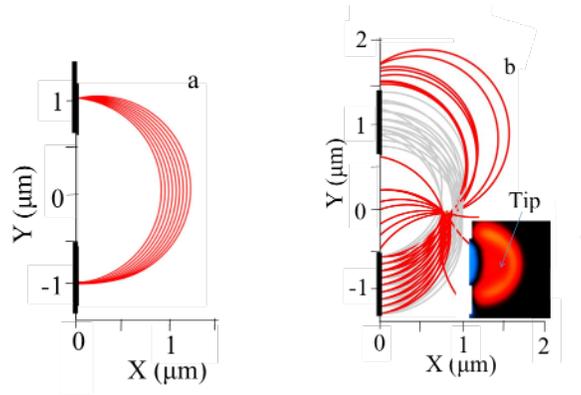

Figure 5: (a) Cyclotron trajectories of electrons injected from a point source in a perpendicular magnetic field. (b) Ray-tracing trajectories for B = 0.130 T and tip position $(0.75, 0)\mu$m. Cyclotron-orbit trajectories are detected by deflecting orbits passing below the tip. The red trajectories are deflected away from the contact [41].

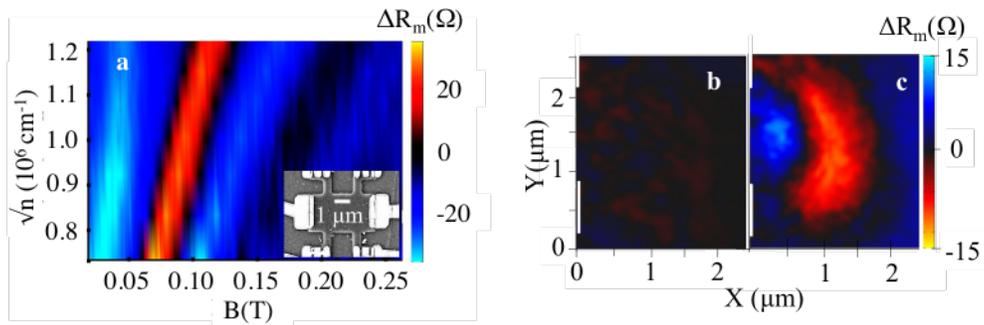

Figure 6: (a) Change in transresistance $DR_m$ vs. magnetic field B and electron density $n^{1/2}$. Inset shows an SEM image of the hBN-graphene-hBN Hall bar sample. (b) Image of electron flow for B = 0; no signal is seen. (c) Image of cyclotron orbits on the first focusing peak in (a). The white lines on the left of the both figures show the point contacts [41].

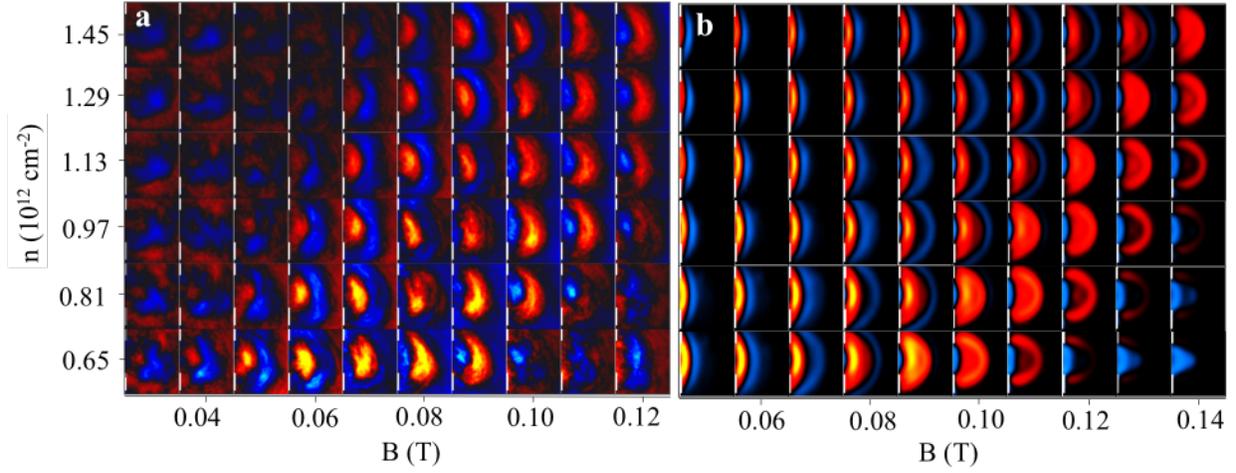

Figure 7: Tiled plots of the transresistance maps as we vary the magnetic field B and electron density n (a) experimental scanning probe microscopy images and (b) ray-tracing simulations. On the first focusing peak in Fig. 6(a), the cyclotron orbits are clearly shown [41].

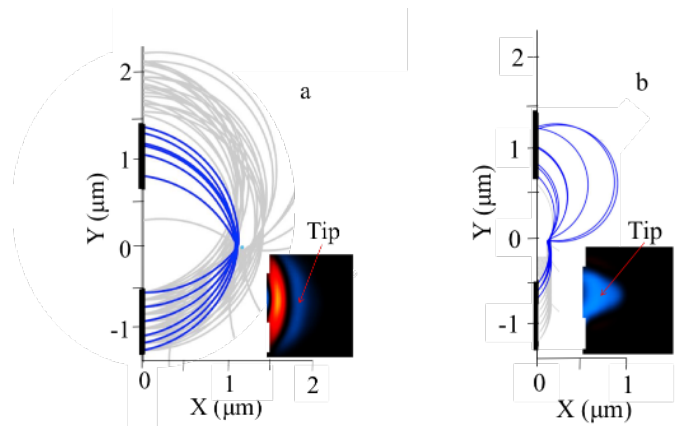

Figure 8: (a) Ray-tracing trajectories for B = 0.09 T and tip position $(1.0, 0)$ $\mu$m. (d) Ray-tracing trajectories for B = 0.140 T and tip position $(0.15, 0)$ $\mu$m. The blue regions correspond to an increase in transmission that are a result of the tip deflecting the electron trajectories (a) into the receiving contact or (b) away from the diffusely scattering edge into the drain contact [41].